\documentclass[twocolumn,superscriptaddress]{revtex4}
\usepackage{amsmath}
\usepackage{graphicx}% Include figure files
\usepackage{dcolumn}% Align table columns on decimal point
\usepackage{bm}% bold math
\def\be{\begin{equation}}
\def\ee{\end{equation}}
\def\bea{\begin{eqnarray}}
\def\eea{\end{eqnarray}}
\newcommand{\eref}[1]{(\ref{#1})}

\newcommand\refeq[1]{Eq.~(\ref{eq:#1})}

\begin{document}

\title{Simulation of grain boundary effects on electronic transport in metals, and detailed causes of scattering}

\author{Baruch Feldman}
\altaffiliation{Current address: Tyndall National Institute, University College Cork, Cork, Ireland}
\email{baruchf@alum.mit.edu}
\affiliation{Process Technology Modeling, Design and Technology Solutions, Technology and Manufacturing Group, Intel Corporation, Santa Clara, CA}
\affiliation{Department of Physics, University of Washington, Seattle, WA}

\author{Seongjun Park} 
\affiliation{Process Technology Modeling, Design and Technology Solutions, Technology and Manufacturing Group, Intel Corporation, Santa Clara, CA}

\author{Michael Haverty}
\affiliation{Process Technology Modeling, Design and Technology Solutions, Technology and Manufacturing Group, Intel Corporation, Santa Clara, CA}

\author{Sadasivan Shankar}
\affiliation{Process Technology Modeling, Design and Technology Solutions, Technology and Manufacturing Group, Intel Corporation, Santa Clara, CA}

\author{Scott T. Dunham}
\affiliation{Department of Physics, University of Washington, Seattle, WA}
\affiliation{Department of Electrical Engineering, University of Washington, Seattle, WA}


\date{\today}

\begin{abstract}
We present first-principles simulations of single grain boundary reflectivity of electrons in noble metals, Cu and Ag.  We examine twin and non-twin grain boundaries using non-equilibrium Green's function and first principles methods. We also investigate the determinants of reflectivity in grain boundaries by modeling atomic vacancies, disorder, and orientation and find that both the change in grain orientation and disorder in the boundary itself contribute significantly to reflectivity.  We find that grain boundary reflectivity may vary widely depending on the grain boundary structure,  consistent with published experimental results.  Finally, we examine the reflectivity from multiple grain boundaries and find that grain boundary reflectivity may depend on neighboring grain boundaries.  This study raises some potential limitations in the independent grain boundary assumptions of the Mayadas-Shatzkes model.
\end{abstract}

\maketitle

\section{Introduction}
  
  As semiconductor technology continues scaling, metal interconnects must scale with minimum feature size (25 nm in the generation currently in development) to connect to nanometer scale devices.  However, in nanoscale metal wires, conductivity can degrade by more than 50\% compared to the bulk metal \cite{SurfScatter,ITRS}.  As a result, both circuit delay times and power consumption may soon be dominated by interconnect \cite{ITRS}.  
  
  The causes of this degradation are believed to be grain boundary, surface roughness, and interface scattering \cite{SurfScatter,ITRS}. As interconnect cross sections decrease, the interactions at the surface or interface with other materials become more important.  Also, grain size is observed to scale roughly with wire thickness \cite{MS,Carreau,Steinhogl}. However, there is limited understanding of the relative importance of these scattering sources on the observed degradation \cite{SurfScatter}, nor is it known to what degree defects and impurities modulate these effects. 
Widely used semi-empirical models of surface \cite{Fuchs,Sondheimer} and grain boundary scattering \cite{MS} and quantum models of surface scattering \cite{surf-theory} have been developed, but there has not been sufficient understanding of grain boundary scattering at the atomic level.  Elsewhere, two of us have studied the effects of wire surface roughness \cite{SurfScatter} and barrier/adhesion/seed layer scattering \cite{Liner} on conduction.  

In this paper, we present atomic-scale modeling and analysis of grain boundary reflectivity in metals and compare these results with experimental data. We also analyze the structure of grain boundaries, investigating in detail the determining factors of scattering. 
The organization of the paper is as follows.  First we present a summary of our method and our results for single-boundary reflectivity.  Next, we give comparisons to published reflectivity measurements, both for multiple- and single-boundary measurements.  Then we give detailed analyses of the causes of grain boundary reflectivity, breaking these into effects of a mismatch between Bloch states in the two crystallites and the effects of disorder and defects.  %Specifically, we investigate the effect of atomic displacements at the boundary and the change in orientation from one grain to another. 
Finally, we raise some concerns regarding the standard model in the literature for relating microscopically calculated reflectivities with macroscopic resistivity, and present our investigations into this model's validity, as well as proposing some suggestions on how the model could be improved. 

\section{Method}

  We use the Non-Equilibrium Green's Function (NEGF) method and the Landauer formalism in this work %[4]. Unlike Boltzmann transport theory, NEGF incorporates both electron propagation and scattering in a single consistent quantum mechanical treatment. The Landauer formulation treats forward and reverse current carriers as in equilibrium with two different potential reservoirs, and therefore can handle contact resistance calculations in ballistic conductors 
  \cite{NEGF}.  We use the code Atomistix\cite{Transiesta} to implement the calculations.  
  In the Landauer formulation\cite{NEGF}, the resistance in perfect metal crystals at absolute zero (ballistic conductors) is contact resistance (the reciprocal of Sharvin conductance).  This is caused by the finite number of transverse modes per unit area with cutoff less than the Fermi energy. The number of modes is equivalent to the cross sectional area of the Fermi surface in a plane normal to the transmission direction:
 \be
 \frac{G_S h}{2 e^2 A} = \frac{T_S}{A} = \frac{M\left(E_F\right)}{A} = \frac{1}{(2 \pi)^2} \int{ \hat{n}_\perp \cdot \hat{z} \; \mathbf{d}^2 k_{||} }
%\frac{T_{c}}{A} =  \frac{M(E_F)}{A} = \frac{1}{(2 \pi)^2} \int \hat{n_\perp} \cdot \hat{z} \; \mathbf{d}^2 k_{||}
\label{eq:cond-ballist}
 \ee
where $G_S$ is Sharvin conductance, $T_S$ is ballistic transmission, $\hat{n_\perp} \propto %\vec{v_k} \equiv 
\vec{\nabla}_k \! E$ is a unit vector normal to the Fermi surface, $\hat{z}$ is the transmission direction, $M(E_F)$ is the number of forward-moving modes with $E = E_F$ , and the integration domain is the set of points on the Fermi surface with $\hat{n_\perp} \cdot \hat{z} > 0$.  %Evaluating \eref{eq:cond-ballist} with a spherical Fermi surface gives $M ( E_F ) / A = 0.1472$ \AA$^{-2}$ for copper, which is in agreement with our simulations (see Table \ref{tab:results} below) and with published results \cite{Xu}. 

  In this paper we investigate the reflection probability $R$ for various structures.  In our notation, the presence of scattering modifies transmission from its ballistic value $T_S$ ( \refeq{cond-ballist} ) to $T \equiv G h / 2e^2 = (1 - R) \: M(E_F)$.  We use computer simulations based on density functional theory\cite{DFT} within the local density approximation (LDA) \cite{PerdewZunger} and NEGF to estimate electron transmission $T$ at 0 K.  
For structural relaxation, we use the total energy pseudopotential method and perform the relaxation within LDA, using augmented wave pseudopotentials with periodic boundary conditions\cite{Payne,Kresse}.  Due to the periodic boundary conditions, we use at least 10 atomic layers for both sides of the grain to avoid the effect of repeating images on the configurations of grain boundaries.  For controlled atomic defect studies due to vacancies and disorder, we do not relax the structures.

\begin{figure}
\includegraphics[scale=0.35]{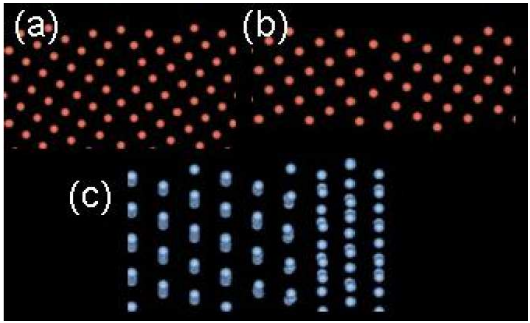}
\caption{ Structure of (a) twin (210) / (120) and (b) twin (320) / (230) boundaries in Cu and (c) non-twin (100) / (110) boundary in Ag.  Transmission is from left to right in all figures. \label{fig:boundaries} }
\end{figure}

  We simulate twin (coincidence site lattice, or CSL) and non-twin grain boundaries in two FCC (face centered cubic) metals, Cu and Ag. We prepare twin boundaries for the two angles with smallest supercells, corresponding to (210) / (120) and (320) / (230), as shown in figures \ref{fig:boundaries}(a) and (b). We also prepare non-twin boundaries for (111) / (110), (110) / (100), and (111) / (100) as demonstrated in figure 
\ref{fig:boundaries}(c). The reflectivity simulation results are summarized in table \ref{tab:results}.

\begin{table}
\caption{\label{tab:results} Summary of simulation results.  All systems are periodic in plane normal to transmission direction. }
%\begin{ruledtabular}
\begin{tabular}{cccc}
System & Relaxed? & $R_\textrm{Cu}$ (\%) & $R_\textrm{Ag}$ (\%) \\
\hline
Twin (210)/(120) & Y &	17 & 12	 \\	
Twin (320)/(230) & Y & 13 & 14 \\
Non-Twin (111)/(110) & Y & 47 & 36 \\
Non-Twin (110)/(100) & Y & - & 46 \\
Non-Twin (111)/(100) & Y & 19 & 16 \\
\\
Vacancy (39.2 \AA$^2$)$^{-1}$ & N & 8 & - \\
Vacancy (19.6 \AA$^2$)$^{-1}$ & N & 16 & - \\
%\\
%Disorder 2 layers & N & 6.9 & - \\
%Disorder 4 layers & N & 22 & - \\
%Disorder 6 layers & N & 24 & - \\
%Disorder 8 layers & N & 27 & - \\
\end{tabular}
\end{table}

\section{Comparison to Experiment \label{sec:validation}}

In this section, we detail the agreement of our calculated grain boundary reflectivity with experimental results.  The difference in length scales (along the transport direction) between those systems that can currently be simulated and actual wires necessarily introduces some uncertainty into the comparison.  Here we compare results despite this theory / measurement gap, but we return to the issue in section \ref{sec:validity-MS}.  

\subsection{Wire Resistivity Measurements}
%\subsection{Low-Temperature Resistivity Measurements}
  In the literature, the Mayadas-Shatzkes (MS) model \cite{MS} is the most widely used analytical model to extract grain boundary reflectivity from experimental measurements of resistivity $\rho$.  The MS model describes metal conductivity as a function of boundary reflectivity $R$ and grain size $D$.  The model's basic assumptions are that grain boundaries are randomly distributed, that all reflections are specular and occur with probability $R$ independent of incoming momentum, and that transport is semiclassical between boundaries. %\footnote{
As we will describe in section \ref{sec:validity-MS}
%later
%, our results from a direct simulation of multiple grain boundaries call into 
several considerations lead us to 
question the validity of these assumptions. %(although the discrepancy with MS predictions may be due solely to minimal bulk scattering over the length scales that can currently be simulated).  
Still, we compare our results for $R$ to measurements of resistivity $\rho$ by using MS, as it is currently the standard model of $\rho$ as a function of microscopic properties. 
%}.  

The MS model agrees with an even simpler model we constructed based only on the assumptions of the Landauer formula \cite{NEGF} and multiple scattering from grain boundaries with average reflection probability $R$.  
%This agreement clarifies the simplicity of the MS model while allowing us to 
This is important because it allows us to extend the model's predictions to more general cases. 
For resistivity of a conductor with both bulk scattering of mean free path $\lambda_b$ and grain boundary scattering, both theories give the resistivity augmentation over bulk as: 
\be
\label{eq:MS}
\frac{\rho}{\rho_b} \approx 1 + \kappa \: \frac{\lambda_b}{D} \: \frac{R}{1-R}
\ee
over most of the range of $R / (1-R)$.  Here 
\[
\kappa = \frac{h A}{  2 M\left(E_F \right) \lambda_b \: \rho_b \; e^2 } \; \approx \; 4/3
\]
in our simplified model and $\kappa \approx 1.39$ in MS \cite{deVries}.  Thus, at room temperature and $D = 45$ nm, a grain boundary reflectivity of 20\% increases resistivity by $\sim$31\% over the bulk value.  

We use the MS model to compare our results for $R$ at 0 K to experimental results at 5 K, in the regime where bulk scattering effects are minimal.
Although Cu is more important for integrated circuits, more experimental data is available on Ag.  The low temperature experiments \cite{deVries} indicate $R \approx 25\%$ for Ag, in comparison to our values of $R \approx 12\%$ for twin boundaries and $R$ from 16\% to 46\% for non-twin boundaries (table \ref{tab:results}).  

%\subsection{Room-Temperature Resistivity Measurements}
A survey of experimental results \cite{GrainRefs} indicates $R$ values for Cu and Ag (computed with MS) in the range of 24\% to 46\%, in good agreement with our predictions.

\subsection{Single-Boundary Resistance Measurements}  
%A review of the literature also identified 
A more direct comparison can be provided by 
a few experiments measuring grain boundary resistance directly in a metal.  Schneider {\em et al.}\cite{Schneider} measured potential difference across single grains in Au (an fcc noble metal like Cu and Ag), and found reflectivity in the range 70\% to 90\%, depending on orientation.  Nakamichi\cite{Nakamichi} measured interface resistivity $\rho_{gb} \equiv R_{gb} A$ (where $R_{gb}$ is interface resistance) in Al for various single grain boundaries.  To compute reflectivity from Nakamichi's results, we estimate the ballistic conductance per unit area $G_c/A$ for Al by \refeq{cond-ballist} assuming a spherical Fermi surface. We then compute reflectivity with 
\be
R = 1 - \frac{A/G_c}{ \left( \rho_{gb}  + A/G_c \right) }.
\ee   
Aggregating Nakamichi's results and analyzing according to this expression, we find most of the measured twin boundary results in the range $R$ = 0\% to 27\% and non-twin results in the range $R$ = 36\% to 51\%.  This range of reflectivity variations agrees well with our findings, particularly given the different materials used.

\section{Determinants of Reflectivity}
 
%In this section, we discuss attempts to categorize and quantify distinct contributions to reflectivity.  
Although real grain boundaries in metals are not necessarily two-dimensional plane defects nor form orthogonal interfaces to the transmission direction, we have isolated the effects and simulated them individually.  

Since a grain boundary is the interface between two crystallites, there are two broad categories of scattering that could occur: 1) scattering caused by the misaligned crystal orientation of the two grains, a category we refer to as {\em orientation effects}; and 2) the atomic structure in the interface itself, which we call {\em atomic position effects}.  We present simulations and analytical arguments to quantify %the contributions of 
these two effects.
  
\subsection{Orientation effects}

%\subsubsection{Identical Bloch states}
   We wish to understand the contribution to scattering from the change in orientation across grains, $R_s$. %A perfect (unrelaxed) twin boundary has mirror symmetry across the boundary, and therefore identical Bloch bases on either side.  Also, an 
An ideal coincidence site lattice (CSL) twin boundary has zero thickness, so the only possible cause of scattering is the abrupt change in orientation.  We therefore consider our results for reflectivity of unrelaxed twin CSLs, $R \sim 15\%$, an estimate of the orientation effect.  

%\subsubsection{Change in Bloch states}
In table \ref{tab:results}, most of the non-twin reflectivities are higher than the twin ones.  This is likely due partly to atomic position effects 
(since the interface is less sharp).  But it may also be due to the difference in Bloch bases in the two grains, particularly if incident states are poorly approximated by states in the other grain.  %One might expect increased reflection due to this difference in transmission across grains.  

We therefore simulate transmission in several different grain orientations. %, with results shown in %Table 
%\ref{tab:orientation}.  
We estimate the orientation effect, due to impedance mismatch between different sets of Bloch states in the two crystallites, by the relative difference 
\[
|\Delta T|/T
\] 
in transmission across the boundary.  (This is by analogy to a simple 1D quantum potential step,  
\[
R = \left(\frac{k-k'}{k+k'} \right)^2, 
\]
where $k$ and $k'$ are the momenta of the states on either side of the step.)  We find that transmission in the $(100)$ and $(111)$ directions are similar, while transmission in the $(110)$ direction is %22\% 
significantly higher. We find transmission for two intermediate orientations $(210)$ and $(320)$ to be between that of $(100)$ and $(110)$.  This shows a large dependence of $T$ on orientation for a perfect crystal, although in a realistic system, we expect that bulk scattering would diminish the orientation effect.  

%\begin{table}
%\caption{ \label{tab:orientation} Cu ballistic transmission {\em vs.} orientation.  $T/A$ values for Ag are generally scaled down from the ones presented in this table by $\left( a_\mathrm{Ag} / a_\mathrm{Cu} \right)^2$ = 1.28.}
%\begin{ruledtabular}
%\begin{tabular}{ccc}
%Orientation & $T/A$ (\AA$^{-2}$) & Normalized to $T(100)/A$ (\%) \\
%\hline
%(100) & 0.1388 & 100 \\
%(111) & 0.1362 & 98 \\
%(110) & 0.1701 & 122 \\
%(210) & 0.1462 & 105 \\
%(320) & 0.1470 & 106 \\
%\end{tabular}
%\end{ruledtabular}
%\end{table}

To confirm this large dependence of transmission $T$ on orientation we compare the simulation results with an analytical estimate of $T$ as a function of orientation in Cu.  The integral in \refeq{cond-ballist} may be evaluated numerically for different directions $z$. The deviations from a spherical Fermi surface then give the effect of grain orientation on ballistic transmission.  %For a non-ballistic conductor, the orientation effect would be muted.  
In the Fermi surfaces of Cu and Ag, the eight (111) directions each contain a ``neck'' that intersects the Brillouin Zone edge.  These necks contribute to \refeq{cond-ballist} for (110) transmission.  
Such a calculation has been carried out by Xu {\em et al.} \cite{Xu}, giving $T(110) / T(100) = 1.07$ and $T(111) \approx T(100)$. This qualitatively agrees with our NEGF-based prediction %in %Table 
%\ref{tab:orientation} 
that the transmission in $(111)$ is similar to $(100)$ and greater in $(110)$ relative to $(100)$.  %The quantitative disagreement between the band theory and NEGF evaluations may be %due the approximation of the Fermi surface as spherical in the analytical Landauer derivation in \refeq{cond-ballist} or 
%due to errors in the NEGF model prediction of the wave functions due to the use of atomic orbitals.

\subsection{Atomic Position Effects}

 We isolate the atomic position effect in the interface by simulating structures with specific structures and extracting the magnitude of the reflectivity, $R_a$. Our initial assumption is that total reflectivity from atomic position in the boundary can be decomposed as a sum of reflectivities from individual defects (individual moved atoms, etc.), 
\be
\label{eq:defect-linearity}
R_a = \sum_i R_{a,i},
\ee 
when the individual reflectivities $R_{a,i}$ are sufficiently small.  
  
\subsubsection{Vacancies}
\begin{figure}
\includegraphics[scale=0.65]{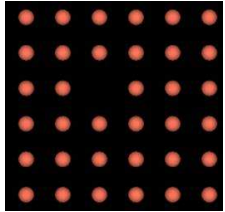}
\caption[Cu lattice with vacancies]{ \label{fig:vacancies} System with plane of vacancies at density %of 1 per 3 unit cells, or 
(39 \AA$^2$)$^{-1}$.  Note that structures are periodic out of the page, but alternate atoms are inequivalent. }
\end{figure}

  We model vacancies in grain boundaries by simulating structures with a single interface containing vacancies as shown in figure \ref{fig:vacancies}. We examine two different vacancy densities, one or two vacancies per 39 \AA$^2$ (3 unit cells) of cross-sectional area. Results are $R$ = 8\% and 16\% (table \ref{tab:results}), scaling linearly with vacancy density consistent with additive reflectivities. The scattering cross section is on the order of magnitude of the area of the missing atoms, as one missing atom per 39 \AA$^2$ corresponds roughly to one vacancy in 12 atoms (3 unit cells) of the interface giving an analytical expectation of $R \sim 1/12 = 8.3\%$, in agreement with the calculations.
  
\subsubsection{Disorder}
\begin{figure}
\includegraphics[scale=0.45]{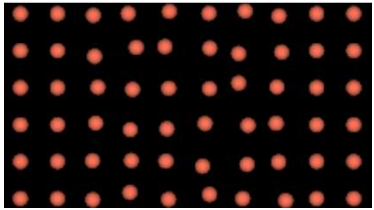}
\caption[Disordered Cu lattice]{ \label{fig:disorder} System with 6 disordered layers (layers 3--8 of those shown). }
\end{figure}

\begin{figure}
\includegraphics[scale=0.35]{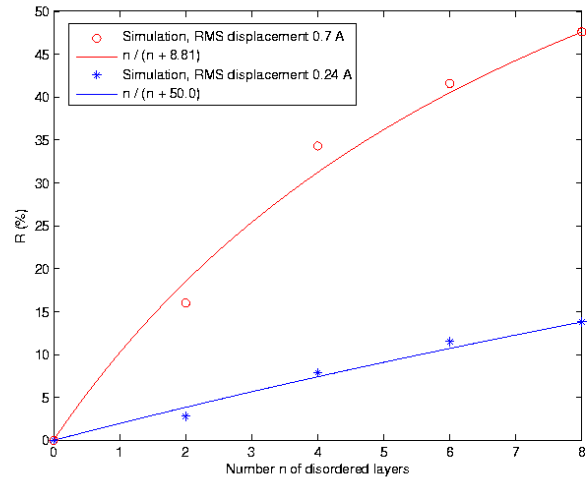}
\caption[Disorder effects on grain boundary conduction]{ \label{fig:results-disorder} Total reflection probability $R$ {\em vs.} number $n$ of disordered layers.  Systems with smaller $n$ have a subset of the disordered region present in systems with larger $n$.  There is significant variability evident due to magnitude of displacement. }
\end{figure}

  To further investigate the effect of atomic position on reflectivity, we simulate layers of disordered Cu atoms as shown in figure \ref{fig:disorder} to isolate the impact of crystalline order on conduction. The disordered atoms are displaced by normally distributed random vectors with root mean square (RMS) magnitude 0.24 \AA~or 0.70 \AA. We change the number $n$ of such layers and expect from theory that $R = n / (n + n_0)$ with a constant $n_0$, since localization length is long compared to our system size\cite{NEGF}.  We summarize the results in figure \ref{fig:results-disorder}. These disordered region simulations give insight into the effect of non-linear boundaries on reflectivity and show that the impact of non-lattice site atomic positions is significant.

  These structures are not the only possible departures from crystalline order, but provide an estimate of the magnitude of the atomic position effect: 
\[
R_a \sim (0.08) \: d_d \; (39 \: \mathrm{\AA}^2),
\] 
with $d_d$ the defect area density in the boundary, and $5\% < R_a < 30\%$, depending on the magnitude of disorder in the boundary. The objective of these estimates is to approximately predict $R_a$, as function of structure. Both estimates reflect a strong dependence of grain boundary reflectivity on atomic position (e.g. gaps, relaxation).

\section{Validity of Mayadas-Shatzkes Model} \label{sec:validity-MS}

We also attempt to check the validity of the Mayadas-Shatzkes (MS) model assumptions using model systems with multiple grain boundaries.  The aim of this investigation is to raise the question: are the assumptions in the MS model, such as the characterization of all grain boundary scattering in a metal by a single reflectivity parameter, tenable?  This undertaking is motivated partly by a lack of consistency in published reflectivities for the same metals.  %, and partly by questions regarding the assumptions themselves.  

The Mayadas-Shatzkes model makes many assumptions, which we summarize here:
\begin{itemize}
\item Transport is semiclassical between grain boundaries.  
\item Grain boundaries are perpendicular to transport.
\item Grain boundaries are translation-invariant along the boundary.  
\item Transmission at grain boundaries can be characterized by a single parameter $R$.
\item All boundaries in a sample are identical (same $R$).  
\end{itemize}
Some of these assumptions are clearly objectionable, and it seems likely that some may affect the final result.  % but many are robust despite being unrealistic.  
For example, consider the assumption of translation-invariant boundaries.  
In the presence of bulk scattering, velocities are randomized within a few mean free paths of the boundary, so the assumption of specular reflection or undeflected transmission may not matter, as long as $R$ matches the average reflection probability.  But if grain size is comparable to mean free path (hard to avoid with wire thickness $\sim 25$ nm $<$ room temperature mean free path of 39 nm), or several boundaries are particularly close together, then the particular states transmitting through one boundary and incident on a second may affect overall transmission.  We simulate just such a system here.  

\begin{figure}
\includegraphics[scale=0.25]{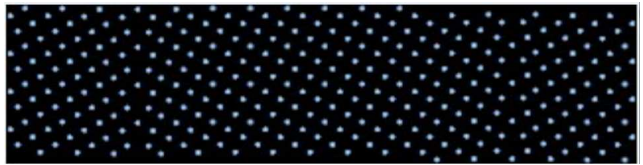}
\caption{System used for validation of MS model. Two twin grain boundaries are separated by an ideal (ballistic) crystalline region. \label{fig:multi-grain}}
\end{figure}

We did simulations with two (210) / (120) twin boundaries separated by from 2.4 nm to 4.6 nm (figure \ref{fig:multi-grain}). Although this differs markedly from an infinite system %with grain size of 45 nm 
as considered in the MS model, the discussion surrounding \refeq{MS} justifies comparing the results using our simplified Landauer transport model for multiple reflection between just two identical boundaries.  This model gives the total reflectivity as 
\be
R_T = 2R / (1+R), 
\label{eq:Land-2}
\ee
where $R$ is the reflectivity of a single boundary\cite{NEGF} (assuming averaging over angles of reflection destroys the coherence in the electron wavefunction between the boundaries).  Using our single-boundary results (table 
\ref{tab:results}), we anticipate $R_T$ = 29\% for Cu and $R_T$ = 21\% for Ag.  

We find our simulation results differ from the smallest to largest grain size as shown in table 
\ref{tab:doublegrain}. We attribute this change in reflectivity to interactions between boundaries.  We also find the simulated $R_T$ is somewhat lower than that predicted by \eref{eq:Land-2}.  This may be explained by the observation that transmission depends strongly on initial momentum in our single-boundary simulations.  The first boundary may act as a filter, letting only those states with highest transmission through to the second boundary.  If a real sample contained a pair of similar boundaries with spacing small compared to the bulk mean free path (like our model systems), one might therefore expect the MS predictions to fail.  The MS model also fails to consider nonspecular reflection and transmission of electrons.  

Still, this explanation would be affected by bulk scattering or variety in boundary type, factors which render the semi-classical MS treatment more tenable.  The failure of MS in the system here could be criticized on the grounds that the simulated system is unrealistic.
Our response is that small grain separation is becoming likelier as wire dimensions shrink, while our neglect of phonon scattering is rendered more realistic by small grain size and the inclusion of disorder in our simulated boundary.  

The ideal test of MS would be a first-principles simulation with multiple, different grain boundaries separated by a more realistic grain size, together with bulk scattering, but this unfortunately is not yet practical.  %The present simulation illustrates a slightly artificial case where the MS assumptions do not hold good, while leaving open the possibility MS could fail in a broader class of cases. 
 The goal of our test has simply been to narrow the theory-measurement gap %as much as possible 
with the computing resources currently available.  
Based on these analytical and computational considerations, we propose that an MS-like model with 1) multiple reflectivity parameters for different incident momenta, 2) a treatment of the statistical likelihood of ballistic propagation between boundaries, and 3) allowance for deflected transport %or nonspecular reflection 
and grain boundary angles 
would capture more relevant physics and probably give more consistency across measurements.  

\begin{table}
\caption{ \label{tab:doublegrain} Total grain reflectivity of two twin grain boundaries with various grain sizes (in units of lattice constant $a_{\rm Cu}=3.6$ \AA, $a_{\rm Ag} = 4.1$ \AA) for Cu and Ag.  %The Landauer prediction would be $R_T = 29\%$ for Cu and $R_T = 21\%$ for Ag for all grain sizes large enough to avoid interaction between boundaries.  
}
\begin{tabular}{ccc}
Grain size/$a$   & $R_T$ for Cu (\%) & $R_T$ for Ag (\%) \\
\hline
6.71 & 23.6 & 19.3 \\
8.94 & 25.8 & 14.7 \\
11.16 & 24.5 & 17.4 \\
\hline
Expected & 29 & 21 \\
\end{tabular}
%\end{ruledtabular}
\end{table}

\section{Discussion}

Our findings show that grain boundary scattering is a significant source of resistivity for sufficiently small grain size.  However, the damascene process can deposit interconnects with average grain size of order the minimum feature size or larger \cite{Carreau,GeissRead,Paik,Steinhogl}, so grain boundary scattering can be reduced but not eliminated.  Grain boundary scattering is still probably more important than the relatively small contribution from surface roughness scattering \cite{SurfScatter},
and even the moderate contribution from barrier layer interface scattering \cite{Liner}.
%Our calculations, in agreement with most measurements of grain boundary reflectivity, suggest that grain boundary scattering is among most important source of conductivity degradation in metal interconnect.  

The damascene process is capable of depositing Cu interconnects with grain size larger than the line thickness.  For example, Geiss and Read\cite{GeissRead} report an average grain diameter of 315 nm for 100 nm damascene Cu lines.  Similarly, Paik {\em et al.}\cite{Paik} measured grain sizes in the range 125 -- 275 nm for $\sim$170 nm line thickness.  Both Carreau {\em et al.}\cite{Carreau} and Steinh\"{o}gl {\em et al.} \cite{Steinhogl} measured somewhat smaller grains for thinner damascene-deposited interconnects, and observed that grain size does indeed scale with thickness for the thinnest wires.  
     
These results suggest that grain sizes on the order of the minimum feature size are readily achievable, and that annealing and overgrowth have the potential to give larger grain sizes.  If we assume that grain size is equal to minimum feature size, \refeq{MS} gives a resistivity augmentation of 42\% for a minimum feature size of 32 nm and average $R = 20\%$, and augmentation of 110\% if the average of $R$ is 40\%.  On the other hand, if a grain size of 3 times the minimum feature size can be achieved (as suggested by the results of Geiss and Read) then resistivity augmentation is only 14\% for $R = 20\%$, and 38\% for $R = 40\%$.

\section{Summary}
  
We have presented the first simulations of reflectivity for relaxed twin and non-twin grain boundaries in Cu and Ag. Our results agree with the experimental reported range of reflectivity \cite{deVries,GrainRefs} and with individual grain boundary measurements \cite{Nakamichi,Schneider}. To gain insight into the mechanisms of grain boundary reflectivity and the impact of the non-planarity of real grain boundaries, we also investigated the effect of vacancies, orientation, and disorder.  We observed that all three contribute significantly to reflectivity. Our predicted dependence of reflectivity on grain boundary type and isolated vacancy, orientation, and disorder effects may explain the wide range of variations in the experimentally data. In probing the utility and extendibility of the Mayadas-Shatzkes model we found that the assumption of a one-parameter reflectivity averaged over all grain boundaries and initial states failed to accurately estimate reflectivity from closely-spaced multiple grain boundaries. Improvements in the analytical models to account for deviations from additivity of multiple boundaries and the impact of grain boundary type and non-planarity are needed.  Based on our studies, it is clear that larger-scale rigorous quantum models are needed to capture more realistic line and grain boundary structures. 

\section{acknowledgement}
We would like to thank Jorge Garcia at Intel for management support of this activity and Profs. Anton Andreev and John Rehr for valuable inputs.

\end{document}